\documentclass[aps, twocolumn, a4paper, superscriptaddress]{revtex4}

\usepackage{graphicx}
\usepackage{float}
\usepackage{color}
\usepackage{amsmath, amssymb}
\usepackage{lmodern}
\usepackage[T1]{fontenc}
\usepackage[utf8]{inputenc}
\usepackage[english]{babel}

\begin{document}

\title{Lotka-Volterra versus May-Leonard formulations of the spatial stochastic Rock-Paper-Scissors model: the missing link}

\author{P.P. Avelino}
\affiliation{Instituto de Astrof\'{\i}sica e Ci\^encias do Espa{\c c}o, Universidade do Porto, CAUP, Rua das Estrelas, PT4150-762 Porto, Portugal}
\affiliation{Departamento de F\'{\i}sica e Astronomia, Faculdade de Ci\^encias, Universidade do Porto, Rua do Campo Alegre 687, PT4169-007 Porto, Portugal}
\author{B.F. de Oliveira}
\affiliation{Departamento de Física, Universidade Estadual de Maringá, Av. Colombo 5790, 87020-900 Maringá, PR, Brazil}
\author{R.S. Trintin}
\affiliation{Instituto de Astrof\'{\i}sica e Ci\^encias do Espa{\c c}o, Universidade do Porto, CAUP, Rua das Estrelas, PT4150-762 Porto, Portugal}
\affiliation{Departamento de Física, Universidade Estadual de Maringá, Av. Colombo 5790, 87020-900 Maringá, PR, Brazil}

\begin{abstract}
The Rock-Paper-Scissors (RPS) model successfully reproduces some of the main features of simple cyclic predator-prey systems with interspecific competition observed in nature. Still, lattice-based simulations of the spatial stochastic RPS model are known to give rise to significantly different results, depending on whether the three state Lotka-Volterra or the four state May-Leonard formulation is employed. This is true independently of the values of the model parameters and of the use of either a von Neumann or a Moore neighborhood. With the objective of reducing the impact of the use of a discrete lattice, in this paper we introduce a simple modification to the standard spatial stochastic RPS model in which the range of the search of the nearest neighbor may be extended up to a maximum euclidean radius $R$. We show that, with this adjustment, the Lotka-Volterra and May-Leonard formulations can be designed to produce similar results, both in terms of dynamical properties and spatial features, by means of an appropriate parameter choice. In particular, we show that this modified spatial stochastic RPS model naturally leads to the emergence of spiral patterns in both its three and four state formulations.
\end{abstract}
    
\maketitle

\section{Introduction \label{sec1}}

Spatial stochastic cyclic predator-prey models have enjoyed considerable success in the modeling of key dynamical features of some biological populations with interspecific competition, including certain populations of {\it E. coli} bacteria \cite{2002-Kerr-N-418-171, 2004-Kirkup-Nature-428-412}  and lizards \cite{1996-Sinervo-Nature-380-240}. One of the simplest models of this type is the so-called spatial stochastic Rock-Paper-Scissors (RPS) model \cite{1996-Sinervo-Nature-380-240, 2002-Kerr-N-418-171, 2004-Kirkup-Nature-428-412,  2007-Reichenbach-N-488-1046, 2007-Reichenbach-PRL-99-238105}, which describes the spatial dynamics of a population of three different species --- usually on square lattice --- subject to reproduction, mobility and (cyclic) predator-prey interactions whose strength is the same for all species (reproduction and mobility) or predator-prey pairs. Although this baseline model has been generalized to include additional species \cite{ 2008-Szabo-PRE-77-011906, 2012-Avelino-PRE-86-036112, 2012-Avelino-PRE-86-031119, 2013-Vukov-PRE-88-022123, 2014-Avelino-PLA-378-393, 2014-Avelino-PRE-89-042710, 2017-Avelino-PLA-381-1014, 2017-Brown-PRE-96-012147, 2019-Park-Chaos-29-051105, 2020-Avelino-PRE-101-062312}, interactions \cite{2004-Szabo-JPAMG-31-2599, 2009-Zhang-PRE-79-062901, 2010-Yang-C-20-2, 2014-Laird-Oikos-123-472, 2014-Rulquin-PRE-89-032133, 2015-Szolnoki-NJP-17-113033, 2016-Szolnoki-PRE-93-062307, 2017-Park-SR-7-7465,   2019-Park-EPL-126-38004, 2021-Bazeia-CSF-151-111255}, and biases \cite{2001-Frean-PRSLB-268-1323, 2009-Berr-PRL-102-048102, 2019-Avelino-PRE-100-042209, 2019-Menezes-EPL-126-18003, 2020-Avelino-PRE-101-062312, 2020-Liao-N-11-6055, 2021-Avelino-EPL-134-48001}, in the present paper we shall only consider its classical three species version.

The classical RPS model has two possible lattice-based formulations, usually referred to as Lotka-Volterra (LV) \cite{1920-Lotka-PNAS-6-410, 1926-Volterra-N-118-558} and May-Leonard (ML) \cite{1975-May-SIAM-29-243} formulations (see \cite{2014-Szolnoki-JRSI-11-0735} for a review). The main difference between the two is the following: in the LV formulation every site on the lattice can be in one of three states (corresponding to the three different species), while in the ML formulation four states are possible at every site (a site may also be empty). In both formulations a von Neumann \cite{1999-Szabo-PRE-60-3776, 2006-Reichenbach-PRE-74-051907, 2007-Reichenbach-N-488-1046, 2008-Reichenbach-PRL-101-058102, 2012-Avelino-PRE-86-036112, 2012-Avelino-PRE-86-031119, 2015-Chen-PRE-92-012819} or a Moore \cite{2019-Menezes-EPL-126-18003, 2019-Mugnaine-EPL-125-58003} neighborhood, composed, respectively, of a central cell and its four or eight adjacent cells, is usually employed.

In the LV formulation of the RPS model, whenever a predator-prey interaction is carried out the predator eliminates its neighboring prey and replaces it with a new individual of the predatory species --- in the LV formulation predation and reproduction happen simultaneously. On the other hand, in the ML formulation, the predator executes a predatory action by eliminating the prey and leaving the neighboring site empty. Hence, unlike in the LV formulation, reproduction and predation correspond to separate actions in the ML formulation. In the three-state LV formulation there is a conservation law for the total number of individuals (always equal to the number of sites), while in the ML four-state formulation the total number of individuals is no longer conserved since the number of empty sites is, in general, time-dependent. 

For small enough mobility rates, both formulations of the spatial stochastic RPS model have been shown to allow for the stable coexistence of all three species in lattice-based simulations. However, the complex spiral patterns, observed in lattice-based simulations of the spatial stochastic RPS model using the ML formulation, are usually absent when the LV formulation is employed. Furthermore, the ML formulation  generally leads to relatively smooth interfaces between different well defined domains --- in the LV formulation the domains are usually fuzzy and do not have well defined boundaries \cite{2008-Peltomaki-PRE-78-031906, 2017-Brown-PRE-96-012147}. This is true independently of the choice of a von Neumann or a Moore neighborhood.

In this paper we shall introduce a simple modification to the standard spatial stochastic RPS model in which the range of the search of the nearest prey/empty site for predation/reproduction may be extended up to a maximum euclidean radius $R$. This change is motivated by the objective of reducing the impact of the use of a discrete grid on the final results, and determining whether or not the observed discrepancies between the results of lattice-based simulations of the standard spatial stochastic RPS model using LV and ML formulations are an artifact associated to the use of a discrete lattice. In particular, we shall investigate if such discrepancies can essentially be eliminated in the modified version of the spatial stochastic RPS model proposed here.

\section{LV and ML formulations of the RPS model \label{sec2}}

In this section we shall describe the LV and ML formulations of the RPS model investigated in the present paper. In both cases, individuals of all three species are distributed on a square lattice with $N^2$ sites and periodic boundary conditions. In the three state LV formulation every site is occupied by a single individual of one of the three-species, while in the four state ML formulation there is also the possibility of a site being empty. The density of individuals of the species $i$ and the density of empty sites shall be denoted by $\rho_i=I_i/N^2$ and $\rho_0 = I_0/N^2$, where $I_i$ and $I_0$ are, respectively, the total number of individuals of the species $i$ and the total number of empty sites (notice that $\rho_0=0$ in the LV formulation).

At every time step, an occupied site and one of its neighboring sites are randomly selected as the active and passive sites, respectively. Then an interaction is randomly selected to be executed by the individual at the active site. In the LV formulation of the RPS model the predator-prey interaction, defined by 
\begin{equation}
i\ (i+1) \to i\ i\,, \nonumber
\end{equation}
with $i=1,...,3$, may be selected (with probability $p$). Here, modular arithmetic is assumed (the integers $i$ and $j$ represent the same species whenever $i=j \, {\rm mod} \, 3$, where $\rm mod$ denotes the modulo operation). On the other hand, in the ML implementation of the RPS model a predator-prey or a reproduction interaction, defined respectively by,
\begin{equation}
i\ (i+1) \to i\ 0\,, \nonumber  \qquad i\ 0 \to ii\,, \nonumber
\end{equation}
may be selected (with probabilities $p$ and $r$, respectively) at every time step. In addition to these, in both LV and ML formulations a mobility interaction, defined by
\begin{equation}
i\ \odot \to \odot\ i\,, \nonumber
\end{equation}
may be selected (with probability $m$) at every time step. Here, $\odot$ represents either an individual of any species or an empty site. If an interaction cannot be executed (for example, if a reproduction interaction is selected and the passive is not an empty site), the above steps are repeated until a possible interaction is performed and the time step completed. A generation time (our time unit) is defined as the time necessary for $N^2$ successive interactions to be completed --- with simultaneous predation and reproduction associated to  a predator-prey interaction in the LV formulation counting as two time steps. Again, notice that the predator-prey interaction has a different meaning in LV and ML formulations of the RPS model (in the LV formulation reproduction is included in the predator-prey interaction while in ML formulation it is defined as a separate interaction). Also, in the LV formulation the total density of individuals is conserved ($\rho_1+\rho_2+\rho_3=1$) which in general does not happen in the ML formulation of the RPS model ---  in the later case, $\rho_0+\rho_1+\rho_2+\rho_3=1$, but $\rho_0$ is no longer required to vanish or to be independent of time.

In most lattice-based studies a von Neumann neighborhood (or a Moore neighborhood), composed of a central cell (the active one) and its four non-diagonal (or its eight) adjacent cells is employed in the definition of the passive individual both in the context of LV and ML formulations of the spatial stochastic RPS model. In this paper we shall consider a simple modification to the standard spatial stochastic RPS model in which the range of the search of the nearest neighbor may be extended up to a maximum euclidean radius $R$: whenever a predator-prey or reproduction interaction is selected the passive individual is chosen, respectively, as the nearest prey or empty site inside an euclidean circle of radius $R$. We will show in the following section that this single modification with respect to the most common choice of neighborhood can be used to greatly attenuate the observed discrepancies between the results obtained using LV and ML formulations of the spatial stochastic RPS model.

\section{Results \label{sec3}}

In this section we shall present and discuss the results of $500^2$ lattice-based numerical simulations of the standard and modified versions of the stochastic RPS model, starting from random initial conditions with $\rho_1=\rho_2=\rho_3=1/3$. Here, we shall consider LV and ML formulations and the following model parameters: $m_{[\rm LV]}= 0.857$, $p_{[\rm LV]} =0.143$ and $m_{[\rm ML]}=0.50$, $p_{\rm [ML]}=0.25$, $r_{[\rm ML]}=0.25$ (LV and ML formulations, respectively); also, $R$ is equal to $R_{[\rm LV]}=13$ and $R_{[\rm ML]}=10$. These values of the parameters were chosen so that LV and ML formulations of the modified spatial stochastic RPS model produce similar quantitative results. A von Neumann neighborhood has been employed when considering the standard version of the lattice-based spatial stochastic RPS model.

In particular, the predation probability is smaller in the LV formulation. This compensates for the fact that predation and reproduction occur simultaneously in the LV formulation, while in the ML formulation they correspond to two separate actions (this also justifies considering equal predation and reproduction probabilities in the ML formulation). In fact, in the ML formulation an individual of the species $i$ may execute a predatory action against an individual of the species $i+1$ leaving behind an empty space, only to see the space again occupied by an individual of the species $i$ as a consequence of a subsequent reproduction interaction. The combined result of these predator-prey and reproduction interactions is null, as if no action had been taken. Hence, a larger predation probability is required in the ML formulation in order to compensate for this effect and to get a similar characteristic dynamical timescale to the one obtained in the context of the LV formulation. 

The other crucial parameter is the radius $R$, which needs to be larger in the ML than in the LV formulation, in order that to match the corresponding characteristic length scales. This difference is related with the fact that if an individual of the species $i$ predates an individual of the species $i+1$ in the ML model, leaving behind an empty site, this site can then be occupied by an individual of the species $i+2$. Hence, for the same value of $R$, the spirals have a larger characteristic length in the ML formulation than in the LV formulation. This effect can be compensated by choosing a larger value of $R$ in the LV formulation. Finally, the mobility probability $m$ was chosen in such a way that $p+m=1$ in the LV formulation and $p+m+r=1$ in the ML formulation.

\begin{figure}[!t]
	\centering
	\includegraphics{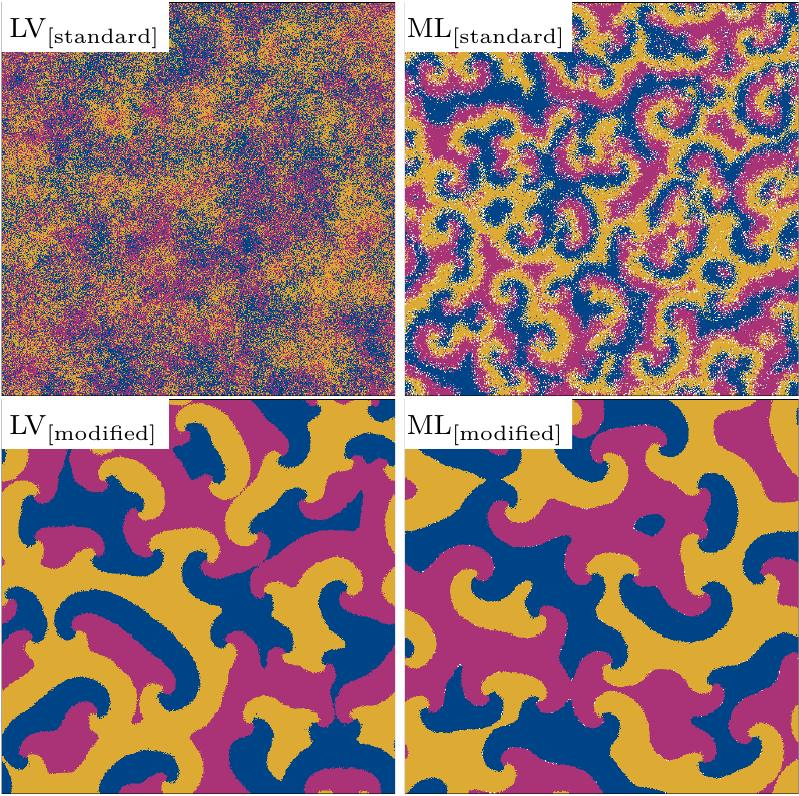}
	\caption{Snapshots of the spatial distribution of the different species on a  $500^2$ lattice at $t=2000$ for LV and ML formulations of the standard and modified versions of the spatial stochastic RPS model. Notice the absence of spiral patterns in the LV formulation of the standard spatial stochastic RPS model and that the significant discrepancies observed for the LV and ML formulations are no longer present in the modified version.}
	\label{fig1}
\end{figure}

Figure \ref{fig1} displays snapshots of the spatial distribution of the different species at $t=2000$. Notice that the spiral patterns are absent in the LV formulation of the standard spatial stochastic RPS model. Also, the significant discrepancies between LV and ML formulations --- which would be present in the standard version  independently of the parameter choice --- essentially disappear in the modified version, for the chosen values of the model parameters. Although the average density of empty sites obtained using the ML formulation (represented in white color),
\begin{eqnarray}
\rho_0^{\rm ML [standard]} &=& 0.102 \pm 0.002\,,\\
\rho_0^{\rm ML [modified]} &=& 0.0018 \pm 0.0001\,, 
\end{eqnarray}
is small both in the standard and modified versions of the spatial stochastic RPS model, it is much smaller in the modified than in the standard version. This happens because in the modified version of the spatial stochastic RPS model, due to the use of an extended neighborhood, the fraction of failed reproduction interactions is much smaller than in the standard one.

\begin{figure}[!t]
	\centering
	\includegraphics{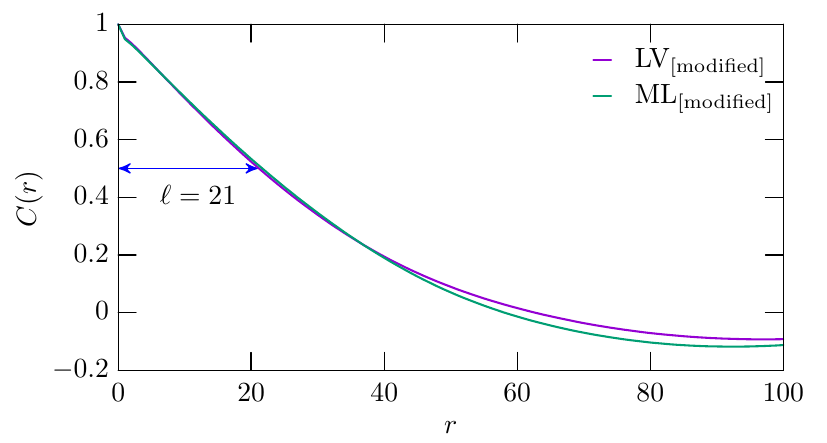}
	\caption{The spatial autocorrelation functions $C(r)$ of the  models considered in the bottom panel of Fig. \ref{fig1} --- the arrow represents the length scale length $\ell$ defined by ($C(r=\ell) = 0.5$). Notice the similarity between the two curves.}
	\label{fig2}
\end{figure}

In order to characterize the size of the structures in the simulations we shall compute a discrete spatial autocorrelation function defined by
\begin{equation}
C(r) =   \sum_{k,l \in S(r)}  \frac{f_{k,l}}{\xi f_{0,0}}\,, 
\end{equation}
where
\begin{eqnarray}
S(r) &=& \left \{ (i,j) \in \mathbb{N}: i + j =\frac{r}{\Delta r} \wedge i \le N \wedge j \le N\right \}\,, \\
f_{k,l} &=&  \sum_{i=1}^N \sum_{j=1}^N \varphi_{i,j} \varphi_{i+k,j+l}\,, \\
\xi &=& {\rm min}[2N - (k+l+1), k+l+1]\,.
\end{eqnarray}
Here, $\Delta r$ is the grid spacing, $\varphi_{i,j} = \phi_{i,j} - \bar \phi$, $\phi_{i,j}$ is equal to the species number of the individual at the position $(i,j)$ on the lattice (or to zero if the site is empty), and $\bar \phi$ represents the mean value of $\phi$. Notice that $C(r)$ is only defined for $r \in \mathbb{N} \wedge r \le 2N$.

\begin{figure}[!t]
	\centering
	\includegraphics{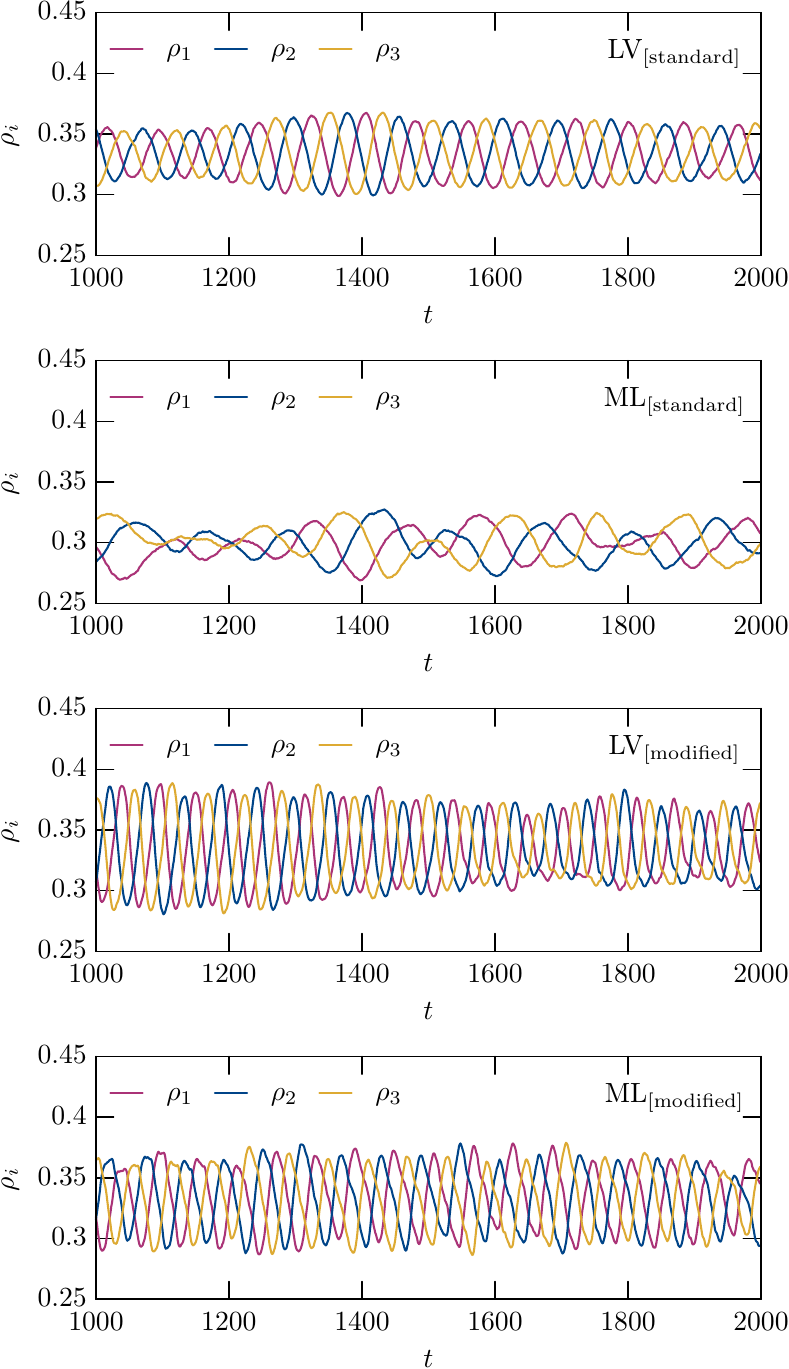}
	\caption{
	The time evolution of the densities $\rho_i$ of the different species over time for the realizations of the spatial stochastic RPS model considered in Fig. \ref{fig1}. Again notice that the observed qualitative and quantitative differences between the results obtained assuming LV and ML formulations of the standard spatial stochastic RPS model are essentially absent in the modified version.}
	\label{fig3}
\end{figure}

Figure \ref{fig2} shows the spatial autocorrelation functions $C(r)$ estimated using simulations of LV and ML formulations of the standard and modified versions of the spatial stochastic RPS model similar to those considered in the bottom panels of Fig. \ref{fig1} --- these functions were obtained considering an average of over $250$ snapshots taken after $6000$ generations. The arrows indicate the approximate value of the characteristic length scale $\ell$ defined as $C(r = \ell) = 0.5$ ($\ell_{[\rm LV]} = 21.2$ and $\ell_{[\rm ML]} = 21.7$, respectively). Although the functions $C(r)$ are only defined for integer values of $r$, we display them as continuous curves for visualization purposes. The similarity between the two curves is a quantitative measure of the agreement between the results obtained using the two formulations (which is clear from the snapshots shown in the two bottom panels of Fig. \ref{fig1}).

Figure \ref{fig3} displays the time evolution of the densities $\rho_i$ of the different species over time for the realizations of the spatial stochastic RPS model considered in Fig. \ref{fig1}. In all four cases there is an oscillatory behaviour, albeit with significantly different properties for LV and ML formulations of the standard version of the spatial stochastic RPS model. These qualitative and quantitative differences are essentially absent in the modified version.

In order to perform a quantitative comparison between temporal series obtained for the LV and ML formulations of the modified version of the spatial stochastic RPS model, let us define the temporal discrete Fourier transform as
\begin{equation}
	\rho_i(f) = \dfrac{1}{N} \displaystyle \sum_{t=0}^{N-1} \rho_i(t)\exp\left[2\pi ift \right]\ ,
	\label{eq3}
\end{equation}
where $\rho_i(t)$ represents the evolution with time of the fractional abundance of a species $i$, and $f$ is the frequency. 

\begin{figure}[!t]
	\centering
	\includegraphics{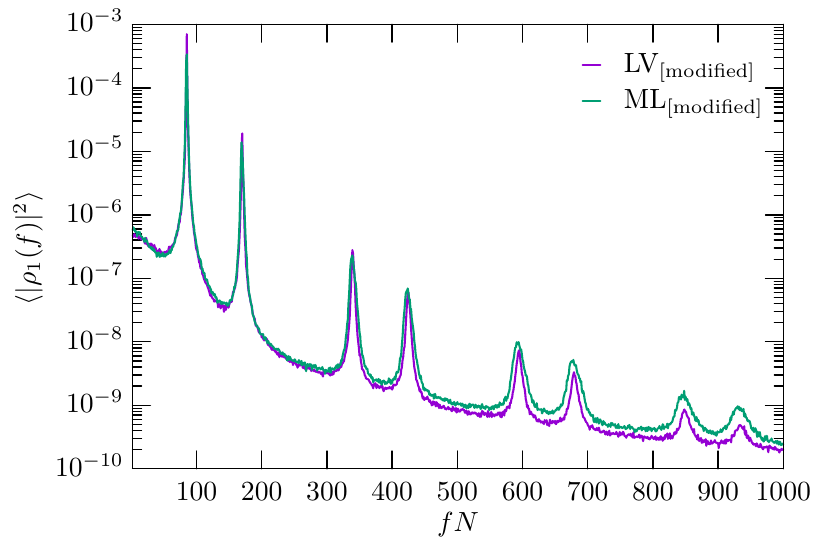}
	\caption{Power spectra of $\rho_1(t)$ obtained using LV and ML formulations of the modified version of the spatial stochastic RPS model (see the two bottom panels of Fig. \ref{fig3}) --- a similar behaviour is found for the species $2$ and $3$. Notice the similarity between the curves and the absence of triplen harmonics.
}
	\label{fig4}
\end{figure}

Figure \ref{fig4} shows the power spectra $\langle |\rho_1|^2\rangle$ estimated using $250$ simulations of LV and ML formulations of the modified version of the spatial stochastic RPS model similar to those used to make the two bottom panels of Fig. \ref{fig3} --- a simulation time span of 6000 generations has been considered, but the first $1000$ have been discarded. Figure \ref{fig4} shows that the visually alike oscillations shown in Fig. \ref{fig3}, have similar power spectra. The first peak of the power spectra, at the first harmonic frequency, occurs at $f \sim 85/N$. Hence, there are about $85$ maximums of $\rho_i$ in a time span of $5000$ generations, or equivalently, the abundance of any of the species $i$ has a maximum roughly every $58.8$ generations. Also notice the absence of triplen harmonics (multiples of the third harmonic). This absence is a consequence of the conservation of the number of individuals (only approximate in the ML formulation) and of the nearly constant displacement (by a third of the quasi period) between the curves of almost identical shape representing the time evolution of the densities $\rho_i$ of the different species shown in Fig. \ref{fig3} (see \cite{CHICCO20111541} for a related discussion in the context of electronics).

\section{Conclusions \label{conc}}

In this paper we investigated the dynamical impact of a simple modification to the lattice-base spatial stochastic RPS model which extends the range of the search for the nearest neighbor up to a maximum euclidean radius $R$. We have shown that, with this modification, the results obtained using LV and ML formulations of the spatial stochastic RPS model --- which significantly differ when a standard neighborhood definition (von Neumann or Moore neighborhood) is used, independently of parameters chosen --- can be brought into agreement by means of an appropriate parameter choice.  We have further shown, that this modified version of the lattice-based spatial stochastic RPS model naturally leads to the emergence of spiral patterns in both LV and ML formulations, which contrasts with their absence in LV lattice-based simulations using the standard neighborhood definition.

These results are in agreement with those obtained using off-lattice simulations where the emergence of spiral patterns may occur in the LV formulation for sufficiently high values of the (conserved) total density of individuals \cite{2018-Avelino-EPL-121-48003}. Hence, the absence of spiral patterns in lattice-based spatial stochastic RPS models employing the LV formulation and a standard neighborhood definition appears to be an artificial effect associated to the fact that the characteristic von Neumman and Moore neighborhood length scale essentially coincides with the lattice spacing. As shown in the present paper, this problem can be resolved by employing an extended neighborhood.

A recent work \cite{2021-Menezes-PRE-103-052216} has found that spiral patterns can appear in the context of spatial stochastic RPS models in which the predation probability is exponentially reduced with the number of preys in the predator’s neighborhood (a circle of radius $R$ centered in the predator). Although, the emergence of spiral patterns in this context was attributed to anti-predator behavior, our results suggest that it is the use of an extended neighborhood which is responsible for relaxing the differences between the results obtained using LV and ML formulations of lattice-based spatial RPS models and for the emergence of spiral patterns in both cases.

\begin{acknowledgments}
P.P.A. acknowledges the support from Fundação para a Ciência e a Tecnologia (FCT) through 
the through the research grants UIDB/04434/2020, UIDP/04434/2020. B.F.O. and R.S.T. thank CAPES - Finance Code 001, Funda\c c\~ao Arauc\'aria, and INCT-FCx (CNPq/FAPESP) for financial and computational support.
\end{acknowledgments}


\end{document}